\documentclass[aps,prd,twocolumn,showpacs,superscriptaddress,groupedaddress,nofootinbib]{revtex4-2}
\usepackage[utf8]{inputenc}
\usepackage{graphicx}  
\usepackage{dcolumn}   
\usepackage{bm}        
\usepackage{amssymb}   
\usepackage[colorlinks=true, linkcolor=blue, citecolor=blue, urlcolor=blue]{hyperref}
\usepackage{xcolor}
\usepackage{array}
\usepackage{booktabs} 
\usepackage{appendix}
\usepackage{ulem}

\usepackage{color}

\begin{document}

\title{Constraints on Halo Gas Profiles from Joint kSZ and Galaxy Clustering Analysis of ACT DR6 and CMASS}

\author{Shaohong Li}
\email{lishh239@mail.sysu.edu.cn}
\affiliation{School of Physics and Astronomy, Sun Yat-Sen University, Zhuhai 519082, China}
\affiliation{CSST Science Center for the Guangdong–Hong Kong–Macau Greater Bay Area, SYSU, Zhuhai 519082, People’s Republic of China}

\author{Yi Zheng}
\email{zhengyi27@mail.sysu.edu.cn}
\affiliation{School of Physics and Astronomy, Sun Yat-Sen University, Zhuhai 519082, China}
\affiliation{CSST Science Center for the Guangdong–Hong Kong–Macau Greater Bay Area, SYSU, Zhuhai 519082, People’s Republic of China}


\date{\today}

\begin{abstract}
{We measure the kinetic Sunyaev–Zel’dovich (kSZ) signal through a joint analysis of the pairwise kSZ effect and galaxy clustering using CMASS galaxies and ACT DR6 maps. This approach breaks degeneracies between the optical depth and nuisance parameters, enabling a reconstruction of the halo optical depth profile as a function of  aperture scale. The kSZ signal reaches its highest signal-to-noise ratio of 7.2 at an aperture radius of $\theta_{\rm AP} = 2$ arcmin, while the full profile rejects the no-kSZ hypothesis at $8.7\sigma$. Applying the same analysis pipeline to the Websky simulation, we find that the observed optical depth profile is somewhat more extended than the simulated one. This difference suggests that baryonic feedback in the real Universe may be stronger and redistribute gas to larger radii more efficiently than modeled in the simulation, although residual systematic effects and modeling uncertainties remain to be further investigated.}
\end{abstract}

\maketitle

\section{Introduction}
\label{sec:intro}

The standard cosmological model, $\Lambda$CDM, has proven remarkably successful in describing the Universe across a wide range of observations, including the cosmic microwave background (CMB), Type Ia supernova, and large-scale structure. Nevertheless, the fundamental nature of its two main components—dark matter and dark energy—remains elusive. The next generation of large-scale structure surveys, particularly those designed to measure weak lensing, such as \textit{LSST/Rubin}~\citep{LSST2012,LSST2019}, \textit{Euclid}~\citep{Amendola2013}, \textit{CSST}~\citep{CSST2026}, and \textit{Roman}~\citep{Spergel2015}, will deliver data with unprecedented precision, sensitive to subtle deviations from $\Lambda$CDM predictions. A critical challenge in harnessing this sensitivity lies in our incomplete understanding of baryonic physics—specifically, the feedback processes that redistribute gas within and beyond halos and thereby alter the matter distribution on nonlinear scales~\citep{vanDaalen2011,Chisari2019}.

Baryonic feedback, driven by processes such as active galactic nuclei (AGN) and supernovae, ejects gas from halo centers and suppresses the matter power spectrum on nonlinear scales. Inaccurate modeling of these effects can introduce significant biases into cosmological parameter constraints, including those on $S_8$ and the total neutrino mass~\citep{Schneider2019,Elbers2025,Huang2019}. Furthermore, baryonic feedback heats a portion of baryons into the diffuse warm-hot intergalactic medium (WHIM) with temperatures in the range of $10^5-10^7$K within and around dark matter halos, thereby exacerbating the problem of the “missing baryons”~\citep{Fukugita2004,Cen2006,Cen1999,Dave1999,Dave2001,Smith2011}.

One promising probe of the redistributed gas is the kinetic Sunyaev–Zel'dovich (kSZ) effect, a secondary CMB anisotropy arising from the scattering of CMB photons off free electrons with coherent bulk motion~\citep{kSZ1970,kSZ1972,kSZ1980,Phillips1995,Birkinshaw1999}. Unlike the thermal SZ (tSZ) effect~\cite{kSZ1972}, which depends on electron temperature and is dominated by hot gas in galaxy clusters, the kSZ effect traces the line-of-sight momentum of all free electrons, making it uniquely sensitive to the distribution of total ionized baryons. This characteristic renders the kSZ effect particularly informative for studying the circumgalactic medium (CGM) and the outskirts of halos. Consequently, measuring the kSZ signal as a function of halo-centric radius provides a valuable probe of baryonic feedback processes.

Recent observational studies have exploited this sensitivity to place constraints on baryonic feedback models. For example, cross-correlations between CMASS galaxies and ACT DR5 data have shown that the level of energy injection in the {IllustrisTNG~\cite{Springel2018} and hydro-simulations in~\cite{Battaglia2010}} is insufficient to push the gas out to the distances from the halo center inferred from observations~\citep{Schaan2021,Amodeo2021}. Using DESI photometric-$z$ and Year 1 samples in combination with ACT DR6 data, \citep{Hadzhiyska2025,Guachalla2025} found evidence that halo gas is more extended than predicted by IllustrisTNG, instead favoring the original Illustris-1 model with stronger baryonic feedback~\cite{Nelson2015}.

Furthermore, by combining DES weak lensing signal~\cite{Sevilla-Noarbe2021} and the kSZ measurements in~\cite{Schaan2021}, \citep{Bigwood2024} constrained the suppression of the nonlinear matter power spectrum, favoring a feedback scenario more extreme than those predicted by most hydrodynamical simulations. Similarly, \citep{McCarthy2025} combined kSZ measurements from BOSS galaxies with \textit{Planck}+ACT maps and galaxy–galaxy lensing, finding that the data prefer stronger feedback than in the fiducial FLAMINGO simulations~\cite{Schaye2023}, whose subgrid models are calibrated to reproduce the local galaxy stellar mass function~\cite{Driver2022} and the gas fractions in low-redshift X-ray-selected groups and clusters~\citep{Kugel2023}. Consistent results were obtained by \citep{Bigwood2025}, who compared DESI Year 1 + ACT kSZ measurements and corresponding galaxy–galaxy lensing signals with a suite of state-of-the-art hydrodynamical simulations, finding that only models with extreme baryonic feedback can reproduce the observations. Incorporating X-ray measurements into a similar analysis, \citep{Siegel2025} reached the same conclusion. 

In several previous studies, observational data have typically been compared with simulation snapshots at fixed redshifts (e.g.,~\citep{Amodeo2021,Hadzhiyska2025}). However, achieving a high signal-to-noise ratio (SNR) in kSZ measurements generally requires integrating over a broad redshift range, which can introduce potential inconsistencies when comparing to single-redshift simulation snapshots. In particular, redshift evolution of baryonic feedback, projection effects from galaxies overlapping along the same line of sight (LOS), and the inverse scaling of halo angular size with angular diameter distance all contribute to light-cone effects. These effects can bias comparisons if not properly accounted for. To address this issue, several recent works have performed like-with-like light-cone comparisons between observations and simulations~\cite{McCarthy2025,Bigwood2025,Siegel2025}. In these analyses, galaxy–galaxy lensing measurements are often used to calibrate the halo mass of the galaxy sample, ensuring a fair comparison between data and simulations. Nevertheless, the inference of the galaxy LOS velocity field differs between observational analyses and simulations, which may introduce additional systematic uncertainties~\cite{Siegel2025}.

To mitigate systematic biases arising from methodological differences in kSZ signal inference, we adopt in this work a unified analysis pipeline that is applied consistently to both observational data and simulated light-cone realizations. We rely on the Websky simulation~\citep{Stein2020}, which provides full-sky CMB maps and corresponding light-cone halo catalogs, to construct mock observations. For kSZ signal extraction, we employ a Fourier-space joint analysis that combines galaxy clustering and the kSZ effect, thereby breaking degeneracies among the galaxy optical depth, the linear growth rate, and nuisance parameters such as the galaxy bias. This framework has been shown to enable high signal-to-noise kSZ detections in~\citep{Li2024}, and has been further extended to cosmological parameter inference in~\citep{Li2025}.

The gas distribution in Websky is modeled using parameterizations calibrated against hydrodynamical simulations~\cite{Battaglia2010}, whose baryonic feedback prescriptions successfully reproduce X-ray and thermal SZ observations~\citep{Battaglia2016,Arnaud2010,Sayers2013,Ade2013,Greco2015}. By varying the aperture radius of the aperture photometry (AP) filter in the kSZ extraction, we measure optical-depth profiles from both observational data and simulations. This enables a direct comparison, allowing us to quantify differences in the gas distribution within halos between the real Universe and the Websky simulation, whose gas distribution models are calibrated from hydro-simulations in~\cite{Battaglia2010}.

This paper is structured as follows. Section~\ref{sec:data} describes the datasets used in this analysis. Section~\ref{sec:method} briefly outlines the methodology. Section~\ref{sec:result} presents and compares results from both observations and simulations. Finally, we discuss our findings and summarize the conclusions in Section~\ref{sec:conclusion}.

\section{Data}
\label{sec:data}
\begin{figure}[t!]
\centering
\includegraphics[width=1\linewidth]{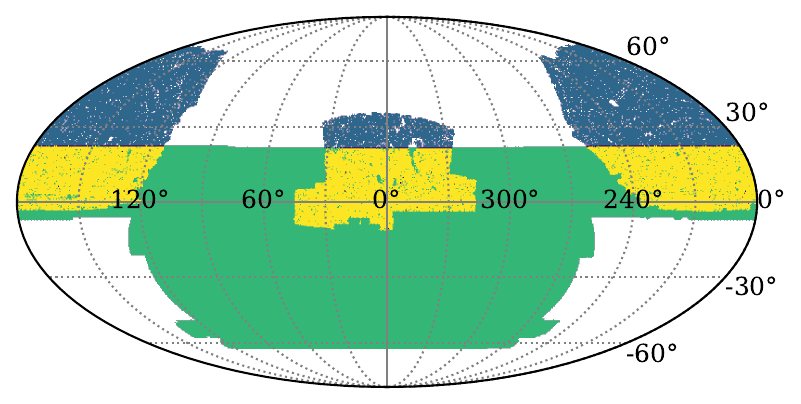}
\includegraphics[width=0.95\linewidth]{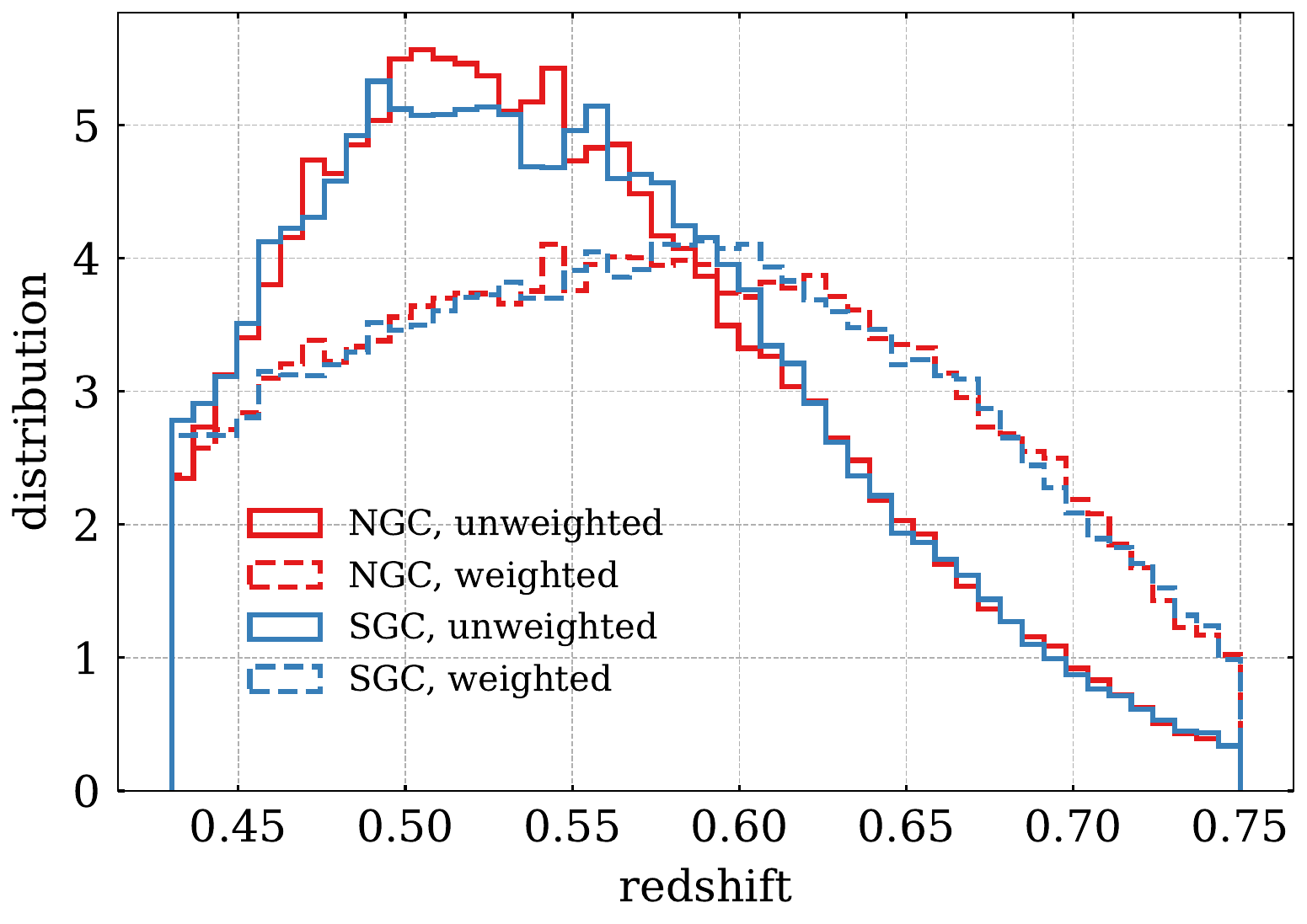}
\caption{\label{fig:distribution}\textit{Top:} Sky coverage of surveys. The yellow region denotes the overlapping area between CMASS and ACT used in this work. The blue, green, and purple regions correspond to the CMASS-only, ACT-only, and overlapping areas that are masked out, respectively. \textit{Bottom:} Redshift distribution of CMASS galaxies. The red and blue solid lines represent the unweighted NGC and SGC samples, while the dashed lines represent the weighted ones. Each profile is normalized to have an integral of 1. The difference between the weighted and unweighted profiles originates primarily from the FKP weight.
}
\end{figure}
\subsection{Atacama Cosmology Telescope map}
\label{subsec:ACT_map}
We extract the kSZ temperature signal from the arcminute-resolution CMB temperature maps of the Atacama Cosmology Telescope (ACT) Data Release 6 (DR6)~\citep{Naess2025}\footnote{\href{https://lambda.gsfc.nasa.gov/product/act/act_dr6.02/act_dr6.02_maps_coadd_get.html}{act-planck\_dr4dr6\_coadd\_AA\_daynight\_f150/f090\_map\_srcfree.fits}}. Specifically, we use the coadded day-night maps at 150 GHz (f150) and 90 GHz (f090) with point sources removed. The effective full width at half maximum (FWHM) is $1.42$ arcmin for the f150 map and $2.07$ arcmin for the f090 map, and both maps have roughly the same median noise level of $14\ \mu\text{K}\cdot\text{arcmin}$. 

To reduce foreground contamination, we apply a mask\footnote{\href{https://lambda.gsfc.nasa.gov/product/act/act_dr6.02/act_dr6.02_maps_ancillary_get.html}{srcsamp\_mask.fits}} that excludes regions with high-contrast areas where point sources have been removed. Additionally, to mitigate edge effects introduced by the aperture photometry (AP) filter, we exclude galaxies located within approximately $10\sqrt{2}$ arcmin of the boundaries of either the CMB map or the survey mask. This scale is set by the maximum aperture photometry (AP) filter radius adopted in this work, $\theta_{\rm AP}=10$ arcmin, which is larger than the value $\theta_{\rm AP}=2$ arcmin used in~\citep{Li2025}.

\subsection{CMASS}

The kSZ signal is maximized at the locations of halos on the CMB map, as traced by galaxies. We adopt the CMASS galaxy sample~\citep{Reid2016} as the tracer population for extracting the kSZ signal. The galaxy catalog used in this work is nearly identical to that in~\citep{Li2025}, with minor differences arising from the larger exclusion regions near survey boundary and masks in this work. The survey footprint covers an area of approximately $4784\,\mathrm{deg}^2$, with galaxies in the redshift range $0.43 < z < 0.75$ and an effective redshift of $z_{\rm eff} = 0.58$. We weight galaxies by $w_{\rm tot}\, w_{\rm FKP}$ when constructing the galaxy number density field, where $w_{\rm tot}$ includes the angular systematics weight, as well as the fiber-collision and redshift-failure nearest-neighbor weights (see Eq.~(50) of~\citep{Reid2016}), and $w_{\rm FKP}$ is the optimal weight that accounts for the redshift-dependent number density~\citep{Feldman1994}. The sky coverage and redshift distributions of the sample are shown in Figure~\ref{fig:distribution}.

\subsection{Websky simulation}

We apply the full analysis pipeline to the Websky simulation~\citep{Stein2020} and compare the results with observational data to investigate the distribution of baryons within and around 
halos. Websky is a widely used suite of high-fidelity simulated sky maps that incorporate multiple cosmological signals, including the primary CMB, tSZ, kSZ, the cosmic infrared background (CIB), radio sources, and weak lensing. Websky generates large-scale structure simulations using the mass-Peak Patch approach~\citep{Stein2019,Bond1996}, an approximate method that combines halo formation modeling with second-order Lagrangian perturbation theory (2LPT) to efficiently produce full-sky light-cone maps for simulating secondary CMB effects. In the Websky kSZ maps, the signal is separated into two components: a dense spherical halo component and a clustered field component. The gas density profiles within halos are modeled using a generalized Navarro–Frenk–White (NFW) profile, with parameters calibrated from hydrodynamical simulations that include AGN feedback (see~\cite{Battaglia2010,Battaglia2016} and other references in~\citep{Stein2020}). {Outside halos (i.e., in the clustered field), the gas distribution is determined by the underlying matter field, and baryonic feedback effects are not explicitly modeled in the kSZ signal on these scales.}

We use the Websky simulation to construct mock samples designed to match both the sky coverage and the redshift distribution of the observational CMASS sample by selecting the most massive halos. To replicate the ACT DR6 f150 and f090 maps, we generate simulated CMB maps by combining multiple microwave components from the Websky simulation. The maps are produced at a HEALPix resolution of $N_{\rm side}=4096$ and include the following components: kSZ, tSZ at 150 GHz (or 90 GHz), CIB at 145 GHz (or 93 GHz), the lensed CMB, and instrumental noise with a level of $14~\mu\text{K}\cdot\text{arcmin}$. The combined maps are then convolved with a Gaussian beam of $\text{FWHM} = 1.42'$ for f150 and $2.07'$ for f090 to match the observational beam profiles.

\section{Methodology}
\label{sec:method}

The methodology of this work is briefly outlined in this section, with further details presented in~\citep{Li2024,Li2025}. We perform the joint analysis of the galaxy density power spectrum and the pairwise kSZ power spectrum to extract the kSZ signal. The theoretical models for these power spectra are derived from nonlinear perturbation theory~\citep{Howlett2019,Qin2025b,Li2025}. To fit the optical depth parameter, which characterizes the kSZ signal, we employ two approaches: separate-fitting analysis and joint-fitting analysis, as applied in~\cite{Li2024} and~\citep{Li2025}, respectively. These are detailed below.

\subsection{Estimator and covariance}
The estimators used in this work are the galaxy power spectrum multipoles ($\hat{P}_{gg}^{\ell=0,2,4}$) and the pairwise density-weighted kSZ power spectrum dipole ($\hat{P}_{\rm kSZ}^{\ell=1}$). The measurement procedure follows that of~\citep{Li2025}, except that we vary the aperture photometry (AP) filter radius over the range $\theta_{\rm AP} = 0.5, 1.0, 1.5, \dots, 10.0$ arcmin in order to measure the optical depth profile.

We estimate the covariance matrices using the jackknife resampling method~\citep{Norberg2009}. Depending on which fitting approach is used for the optical depth, we separately derive the covariance matrices for the galaxy multipoles and the kSZ dipole, as well as their joint covariance matrix. For the jackknife estimation, we divide the data into $N_{\rm JK}=1024$ subsamples. 

The covariance matrix is computed as
\begin{equation}
\label{eq:Cij}
    C_{\rm JK} = \frac{N_{\rm JK}-1}{N_{\rm JK}} \sum_{k=1}^{N_{\rm JK}} (y_i^k - \bar{y}_i)(y_j^k - \bar{y}_j)\,,
\end{equation}
where $y_i^k$ denotes the $i$th data vector element (either $P_{gg}^{\ell=0,2,4}$, $P_{\rm kSZ}^{\ell=1}$, or their joint combination) from the $k$th jackknife subsample, and $\bar{y}_i$ is the mean over all subsamples. The robustness of the jackknife methodology is verified in the appendix E of~\citep{Li2025}.

The inverse covariance matrix is obtained by rescaling $C_{\rm JK}^{-1}$ with the Hartlap factor~\citep{Hartlap2007}:
\begin{equation}
C^{-1} = \frac{N_{\rm JK} - N_{\rm bin} - 2}{N_{\rm JK} - 1} \, C_{\rm JK}^{-1}\,,
\end{equation}
where $N_{\rm bin}$ is the number of data bins: $45$ for the galaxy multipoles, $15$ for the kSZ dipole, and $60$ for their joint combination.

\begin{figure*}[t!]
\centering
\includegraphics[width=0.45\linewidth]{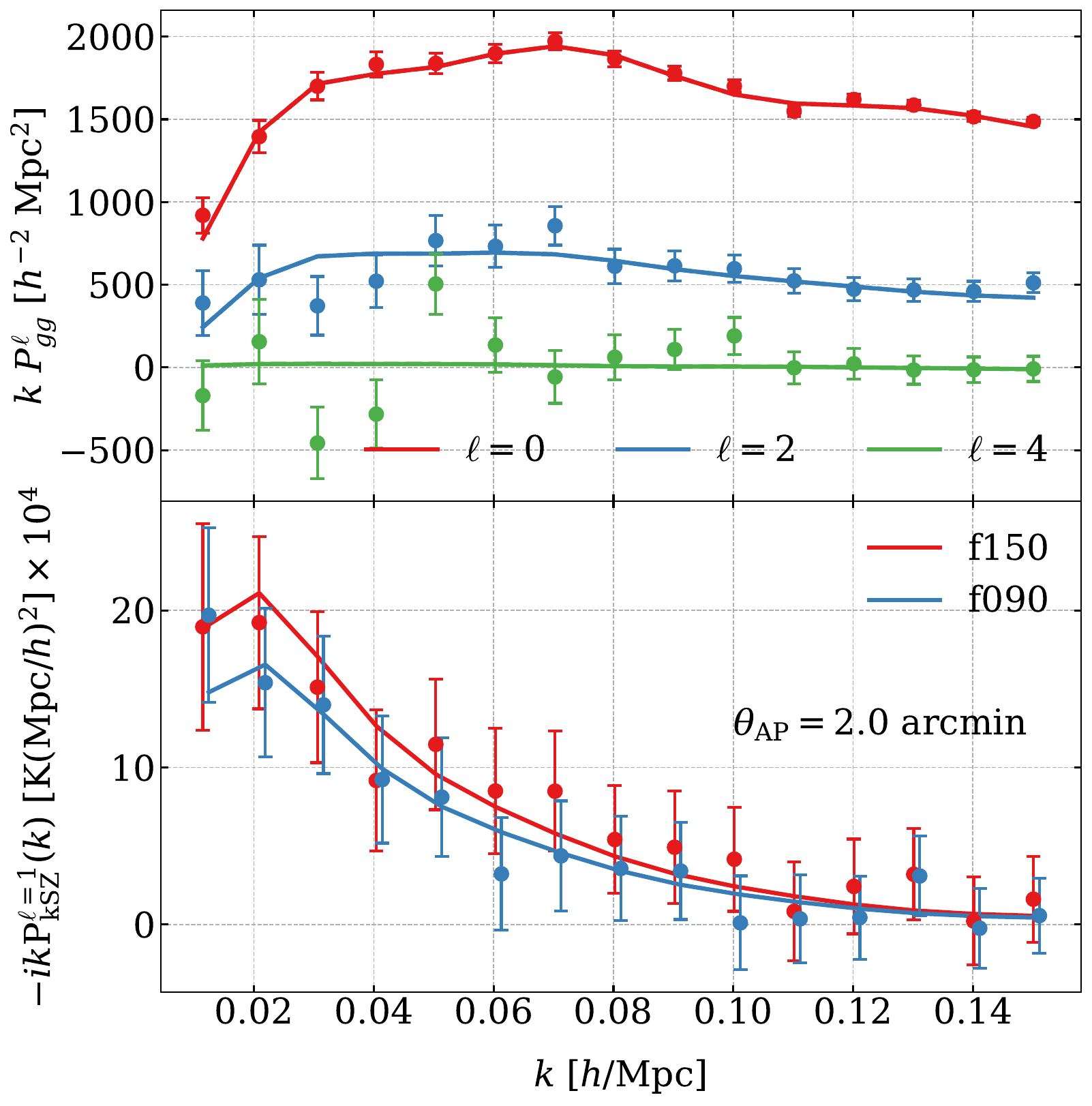}
\includegraphics[width=0.45\linewidth]{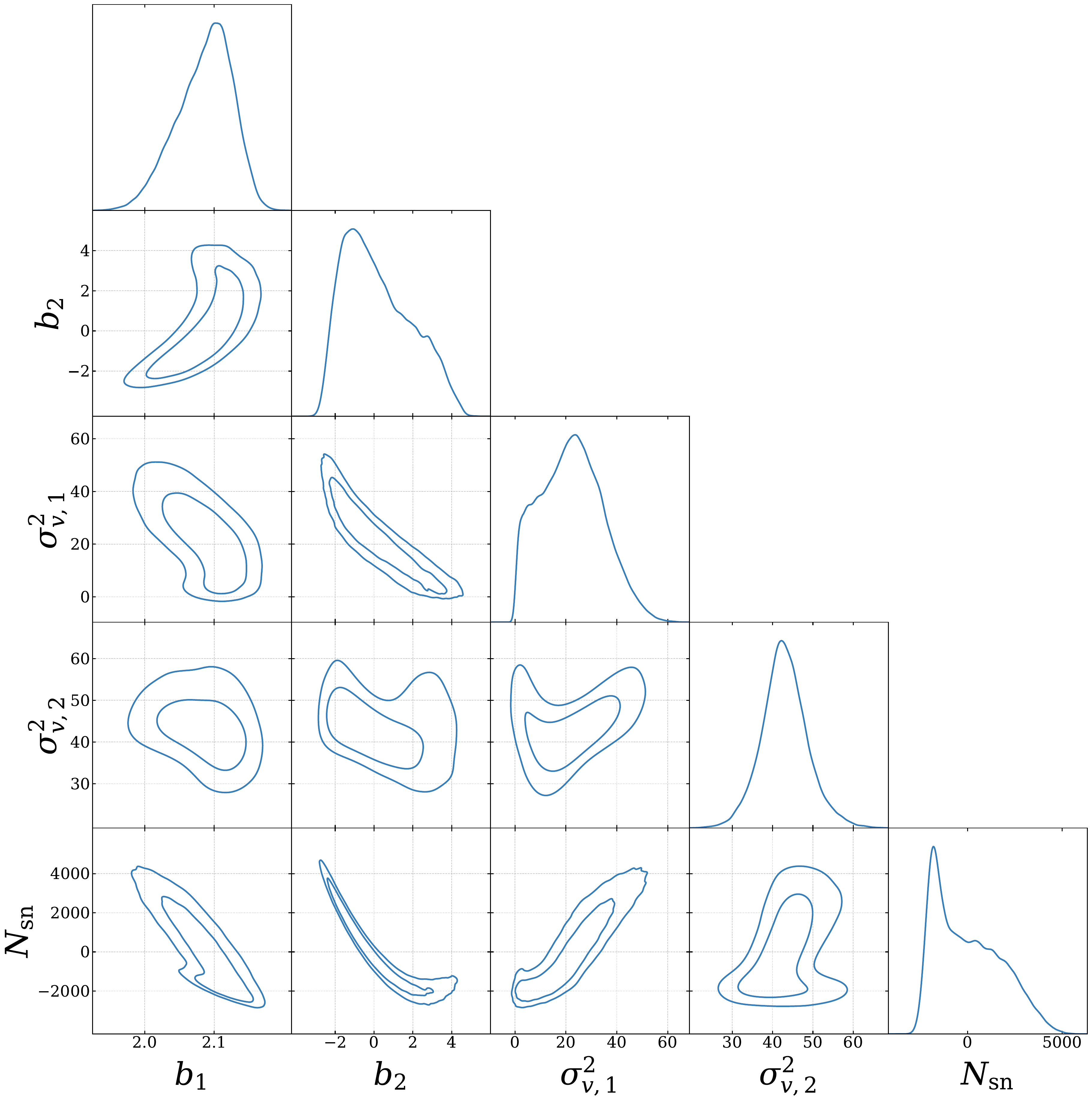}
\caption{\label{fig:Pgg} The fitting results of the separate-fitting analysis. \textit{Upper left:} the galaxy power spectrum multipoles. The lines indicating the best-fit model obtained from fitting the galaxy multipoles alone. \textit{Lower left:} the kSZ power spectrum dipole for an aperture radius of $\theta_{\rm AP}=2.0$ arcmin. The red and blue lines correspond to the best-fit results for the f150 and f090 frequency bands, respectively. \textit{Right:} the posterior distributions of the parameters from the galaxy power spectrum multipole fit.
}
\end{figure*}

\subsection{Theory}
\label{subsec:theory}
We adopt the theoretical model presented in~\citep{Li2025}, which is derived within the framework of nonlinear perturbation theory~\citep{Howlett2019,Qin2025b}. The common free parameters for the power spectrum multipoles include three cosmological parameters: the linear growth rate $f$, and the two Alcock–Paczyński parameters $\alpha_\parallel$ and $\alpha_\perp$~\citep{Alcock1979}, as well as five nuisance parameters: the linear galaxy bias $b_1$, the second-order bias $b_2$, two velocity dispersion parameters $\sigma^2_{v,1}$ and $\sigma^2_{v,2}$, and a residual shot-noise term $N_{\rm sn}$. In addition, the kSZ dipole $P_{\rm kSZ}^{\ell=1}$ introduces an extra free parameter, the mean optical depth $\bar{\tau}$, which encodes astrophysical information related to the baryon distribution.

\subsection{Fitting strategy}

In general, a joint analysis of the galaxy multipoles and the kSZ dipole effectively breaks the degeneracy between $\bar{\tau}$ and the nuisance parameters, particularly the linear bias $b_1$. We separately follow and compare the fitting strategies in~\cite{Li2024} and~\cite{Li2025} to constrain the galaxy optical depth $\bar{\tau}$ in this work.  These two approaches are described below.

In the first separate-fitting analysis approach, we perform fitting analyses of the galaxy power spectrum and the kSZ power spectrum separately. The galaxy power spectrum is used to constrain the nuisance parameters while the cosmological parameters are held fixed {to \textit{Planck18} results~\citep{PLANK2018}}. Once the best fitted values of nuisance parameters are determined, the kSZ dipole depends only on $\bar{\tau}$, which scales its amplitude. Following the method outlined in Section 5.1 of~\citep{Li2024}, we then fit $\bar{\tau}$ to the measured kSZ dipole, obtaining both its uncertainty and the corresponding SNR. This approach not only ensures that the nuisance parameters can be well constrained, but also yields a high SNR for the kSZ signal. Given that our main goal is to study the halo baryon distribution via the $\bar{\tau}$ profiles, where high SNR is essential, we present the main results of this work using this approach.

In the second approach, we carry out a joint-fitting analysis of the galaxy multipoles and the kSZ dipole, fitting all parameters—including cosmological, nuisance, and astrophysical ($\bar{\tau}$) ones—simultaneously via the Markov Chain Monte Carlo (MCMC) method. Implementation details are provided in~\citep{Li2025}. This approach properly accounts for uncertainties from both cosmological and nuisance parameters, ensuring an accurate assessment of the optical depth uncertainty while simultaneously constraining cosmological parameters~\citep{Li2025}. By demonstrating the consistency of the inferred $\bar{\tau}$ profiles between the two approaches in section~\ref{subsec:joint_analysis}, we show that fixing cosmological parameters and using best-fit values inferred from the density power spectrum multipoles does not bias the resultant $\bar{\tau}$ profiles. Conversely, this also demonstrates the potential to jointly constrain cosmology and baryon distribution through a combined analysis of galaxy clustering and the kSZ effect with improved observational data in the future.

\begin{figure*}[t!]
\centering
\includegraphics[width=0.45\linewidth]{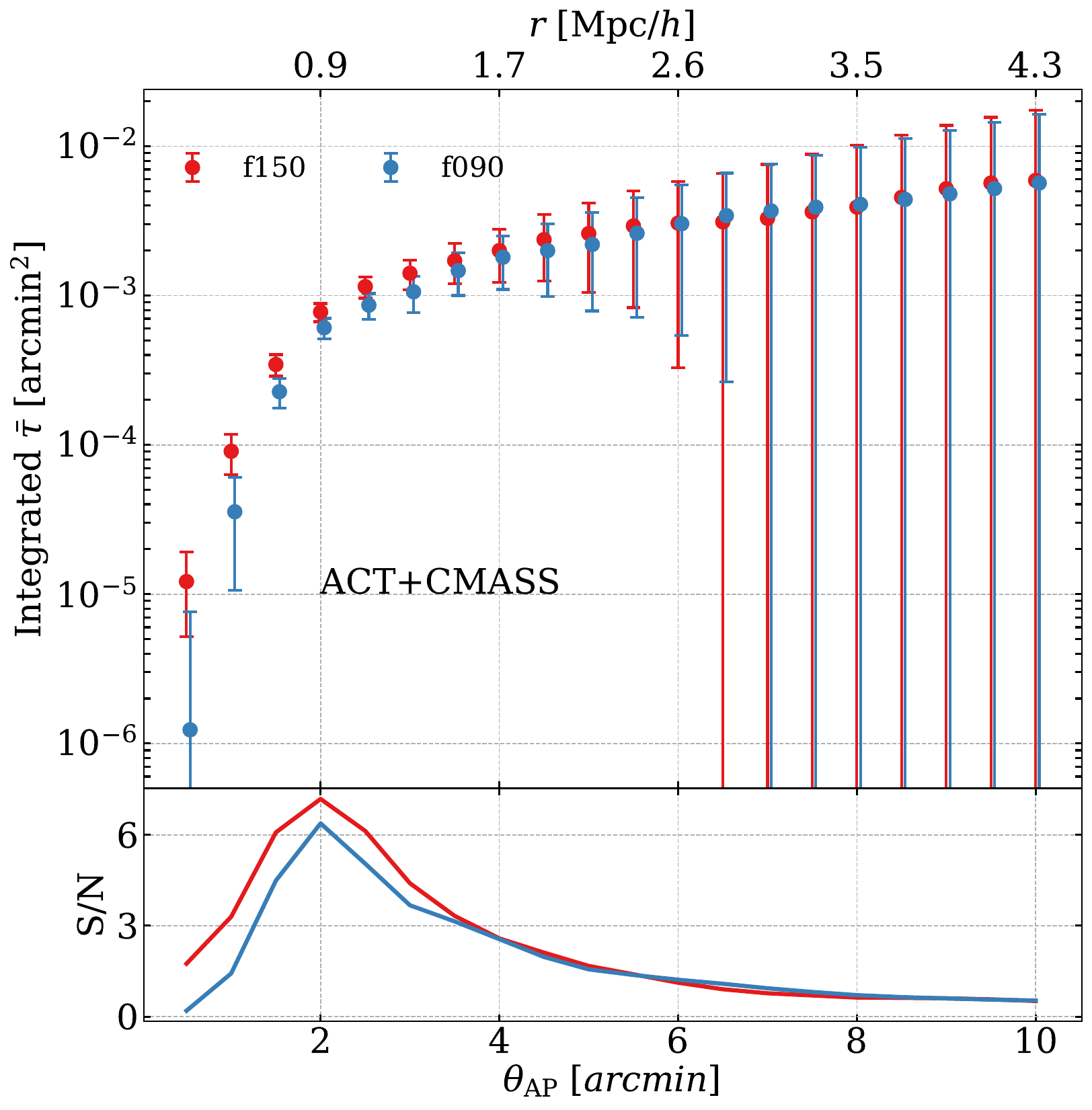}
\includegraphics[width=0.47\linewidth]{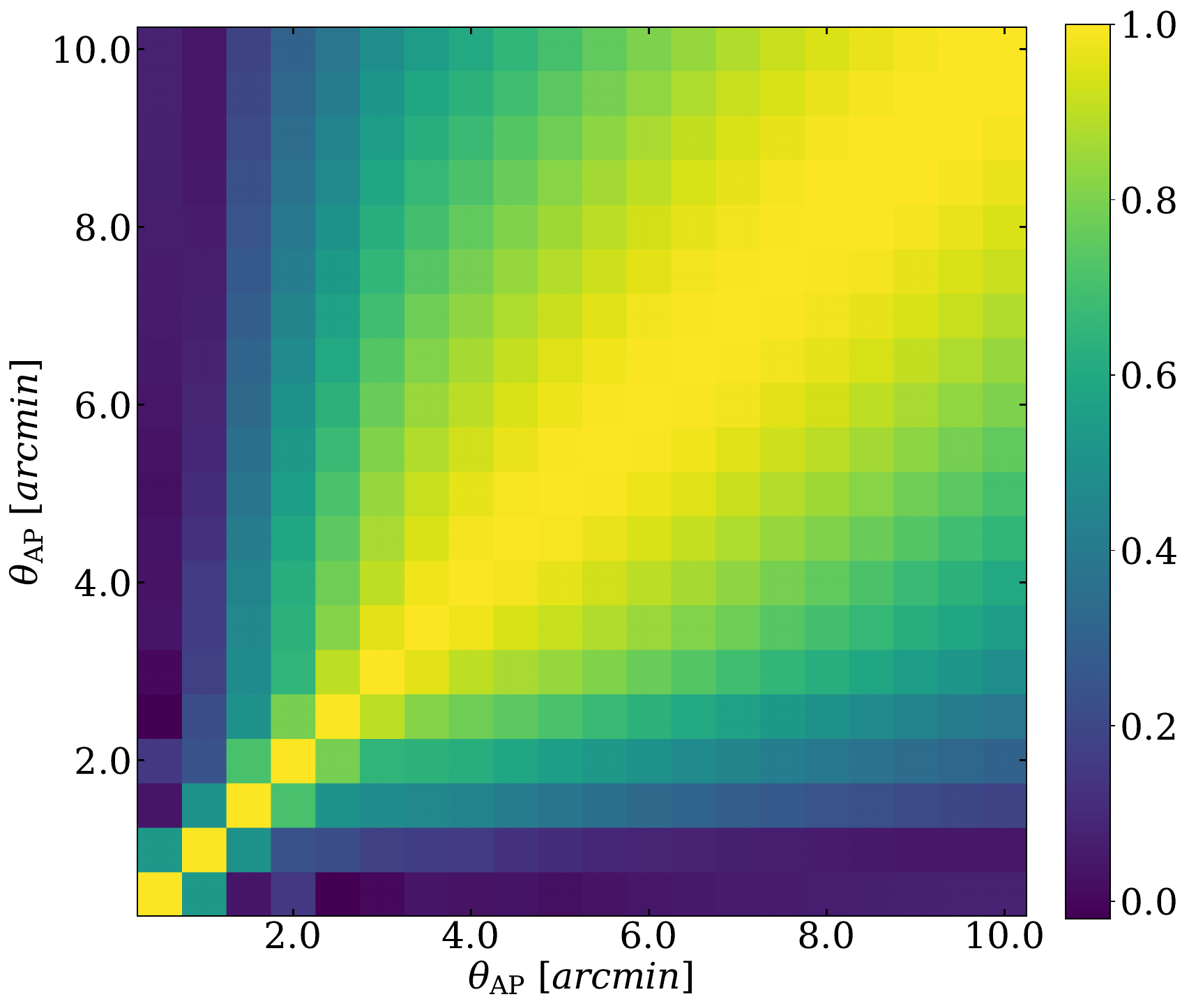}
\caption{\label{fig:tau_profile} \textit{Left:} the integrated optical depth profile, defined as ${\rm Integrated} \ \bar{\tau} = \pi \theta_{\rm AP}^2 \, \bar{\tau}$, with red and blue lines corresponding to the f150 and f090 frequency bands, respectively. The lower subpanel displays the corresponding SNR, which peak at $\theta_{\rm AP} = 2.0$ arcmin with values of 7.2 for f150 and 6.4 for f090. The top axis indicates the equivalent radius in comoving coordinates. \textit{Right:} the correlation coefficient matrix for the $\bar{\tau}$ measurements.}
\end{figure*}
\section{Results}
\label{sec:result}

The results of this work are presented in this section. We first measure the integrated $\bar{\tau}$ profiles of ACT+CMASS using the separate-fitting analysis and compare the results with those from the Websky simulation to study the baryon distribution within and around halos. We then perform the joint-fitting analysis and compare its results with those obtained from the separate-fitting analysis.

\begin{figure*}[t!]
\centering
\includegraphics[width=1.0\linewidth]{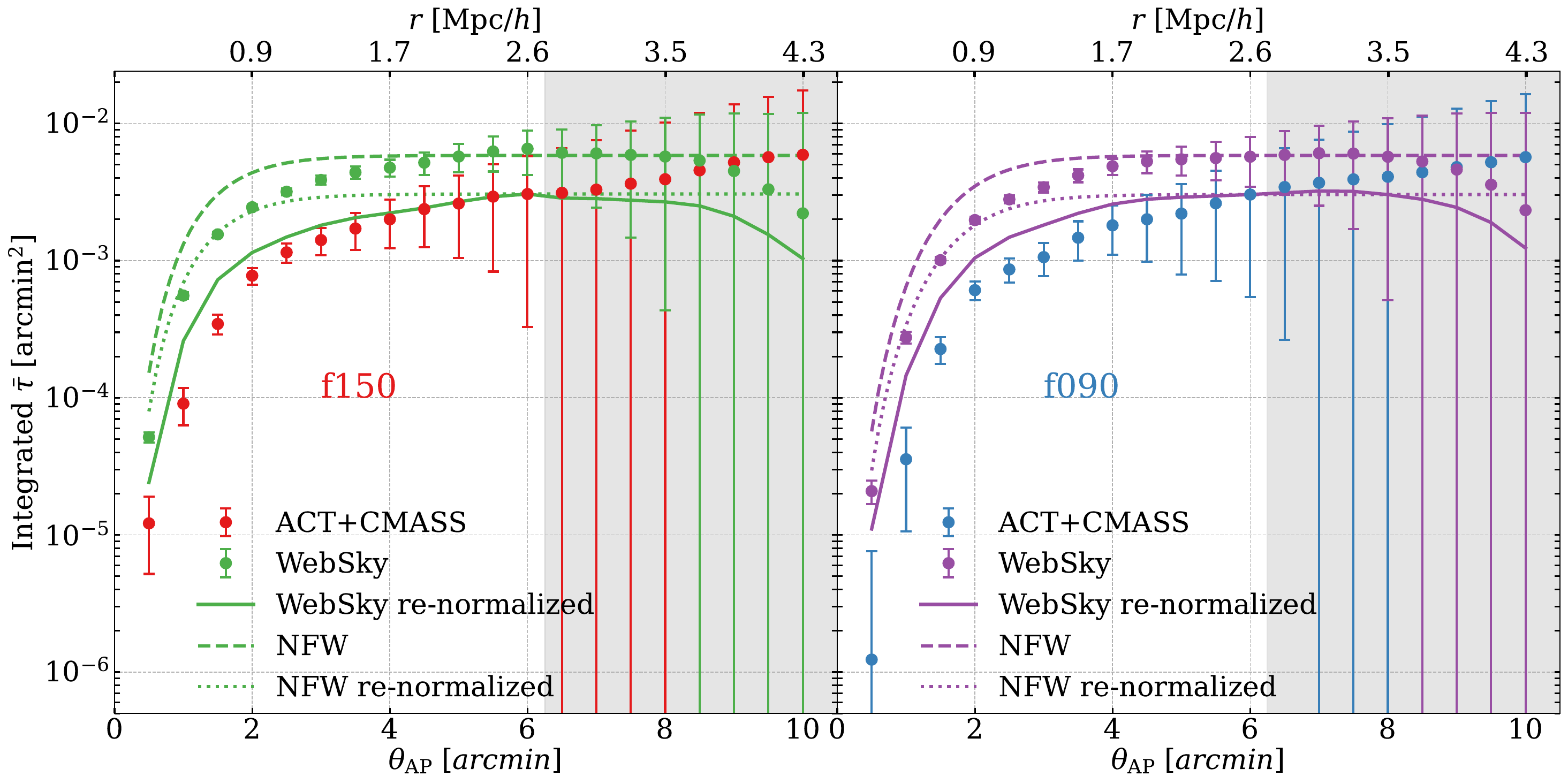}
\caption{\label{fig:tau_profile_all} \textit{Left:} Results for the f150 band. Shown are the integrated $\bar{\tau}$ profiles from ACT+CMASS (red points with error bars), the Websky simulation (green points with error bars), and the NFW model prediction (green dashed line). The green solid and dotted lines correspond to the Websky and NFW profiles, respectively, renormalized to match the ACT+CMASS measurement at $\theta_{\rm AP} = 6$ arcmin. \textit{Right:} Same as the left, but for the f090 band.}
\end{figure*}
\subsection{Separate-fitting analysis}

\subsubsection{Integrated $\bar{\tau}$ profiles.}

We perform the separate-fitting analysis in this subsection. The measured galaxy power spectrum multipoles are presented in the top left panel of Figure~\ref{fig:Pgg}, along with the best-fit multipole models. The posterior distributions of the fitted nuisance parameters are shown in the right panel. In the lower left panel, we present the kSZ dipoles from two ACT maps at different frequencies, with the AP filter radius $\theta_{\rm AP}=2.0$ arcmin. Overall, the amplitude of the kSZ dipole in the f150 band at $\theta_{\rm AP}=2.0$ arcmin is larger than that in the f090 band, primarily due to the smaller FWHM of the f150 map.

Then we vary the aperture size $\theta_{\rm AP}$ and estimate $\bar{\tau}$ as a function of scale, yielding the $\bar{\tau}$ profile. Consistent with~\citep{Schaan2021,Hadzhiyska2024}, we consider the integrated optical depth, defined as $\textit{Integrated}\,\,\bar{\tau} = \pi \theta_{\rm AP}^2 \, \bar{\tau}$. Ignoring additional structures along the LOS and other halos within the aperture, the integrated $\bar{\tau}$ describes the differential number of free electrons of the target halo within the inner circle of the aperture relative to that in the outer annulus. When the aperture radius becomes sufficiently large to reach a region where the gas density is approximately constant, the integrated $\bar{\tau}$ measures the excess free electrons of the halo relative to the field level. At this point, it no longer increases with larger aperture radius and instead flattens. 

The results are shown in the left panel of Figure~\ref{fig:tau_profile}. Profiles at two frequencies overlap at large aperture radius, while the instrumental beam affects the profile at small $\theta_{\rm AP}$. Owing to its larger FWHM, the $\bar{\tau}$ profile of f090 is more strongly suppressed on small scales than that of f150.  The corresponding highest SNRs are 7.2 for f150 and 6.4 for f090, both achieved at $\theta_{\rm AP}=2.0$ arcmin. 

The covariance matrix among $\bar{\tau}$ measurements at different aperture scales is estimated following the method of~\citep{Sugiyama2018}:
\begin{eqnarray}
\label{eq:Cij_tau}
&&{\rm Cov}(\bar{\tau}(\theta_{\rm AP}),\bar{\tau}(\theta_{\rm AP}^\prime)) =\sum_{i,j }\sum_{k,l }{\rm Cov}\left(\hat{P}_i(\theta_{\rm AP}),\hat{P}_k(\theta_{\rm AP}^\prime) \right) \nonumber \\
&&\times\frac{C_{ij}^{-1}(\theta_{\rm AP})\bar{P}_j}{\sum_{k,j}\bar{P}_i C_{ij}^{-1}(\theta_{\rm AP})\bar{P}_j} \quad \times \frac{C_{kl}^{-1}(\theta_{\rm AP}^\prime)\bar{P}_l}{\sum_{k,j}\bar{P}_i C_{ij}^{-1}(\theta_{\rm AP}^\prime)\bar{P}_j}\,,
\end{eqnarray}
where ${\rm Cov}\left(\hat{P}_i(\theta_{\rm AP}),\hat{P}_k(\theta_{\rm AP}^\prime) \right)$ denotes the covariance of the kSZ dipole $P_{\rm kSZ}^{\ell=1}$ between different $\theta_{\rm AP}$, and $\bar{P}_i$ is the theoretical model for $P_{\rm kSZ}^{\ell=1}$ evaluated at $\bar{\tau}=1$.

The correlation coefficient matrix derived from this covariance is shown in the right panel of Figure~\ref{fig:tau_profile}. Strong correlations are observed among the off-diagonal terms, particularly at large $\theta_{\rm AP}$. This is expected, as the integrated optical depth at larger apertures encompasses all the information from smaller scales.

\subsubsection{Comparison with Websky simulation}

\begin{table}[t!]

\centering
\begin{tabular}{c|c|c|c}
\hline
Frequency & $\rm SNR_{null}$ & $\rm SNR_{Websky}$ & $\rm SNR_{NFW}$ \\
\hline
f150 & 8.7 & 7.4 & 27.5  \\
f090 & 7.3 & 6.9 & 21.3  \\
\hline
\end{tabular}
\caption{\label{tab:SN} {Detection SNR.}}
\end{table}

The integrated $\bar{\tau}$ profiles of ACT+CMASS are shown in Figure~\ref{fig:tau_profile_all} again for comparison with those from the Websky simulation. For the f150 channel, the profile appears to flatten at $\theta_{\rm AP} \sim 6$ arcmin, and shows a mild increase at larger apertures. In contrast, for the f090 channel, the integrated $\bar{\tau}$ continues to increase up to $\theta_{\rm AP} = 10$ arcmin. Given the large measurement uncertainties at $\theta_{\rm AP} > 6$ arcmin, it is reasonable to conclude that the integrated $\bar{\tau}$ effectively flattens at $\theta_{\rm AP} \gtrsim 6$ arcmin. This scale corresponds to $\sim 2.6\,{\rm cMpc}/h$ at $z=0.58$, which is approximately $2$–$3$ times larger than the typical halo virial radius. This result is consistent with previous studies~\cite{Schaan2021,Amodeo2021,Hadzhiyska2025,Guachalla2025}.
\begin{figure*}[t!]
\centering
\includegraphics[width=1.0\linewidth]{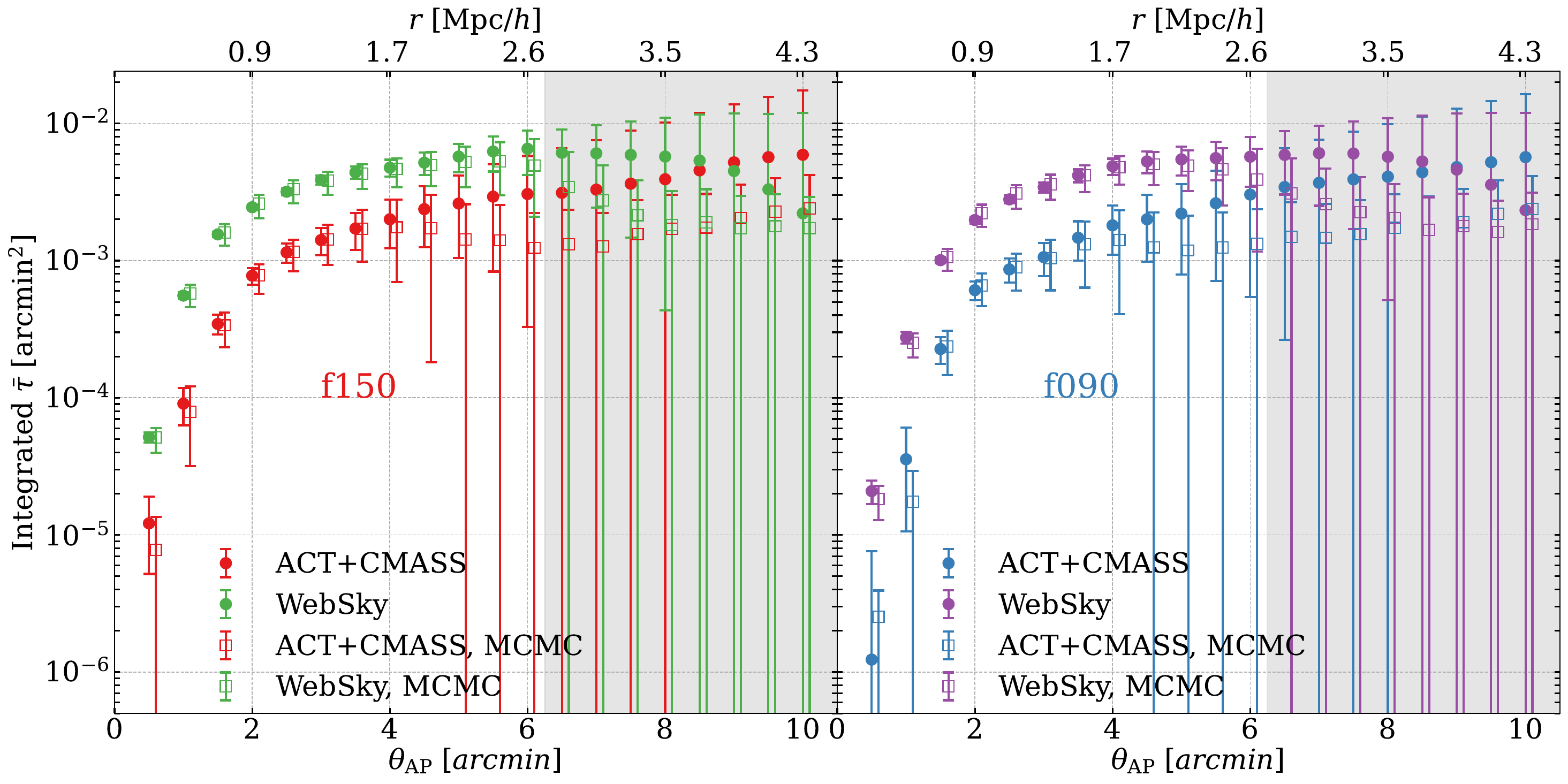}
\caption{\label{fig:tau_profile_mcmc} The points with error bars are the same as those shown in Figure~\ref{fig:tau_profile}. The open squares represent the $\bar{\tau}$ values obtained from the joint-fitting analysis via MCMC.}
\end{figure*}

We then use the Websky simulation to mock the ACT+CMASS analysis, with results also shown in Figure~\ref{fig:tau_profile_all}. We generally observe an offset of the mock $\bar{\tau}$ profiles above the observed ones, in particular at $\theta_{\rm AP}$ where the measurements are relatively significant. As discussed in~\citep{Li2025}, this may be due to at least two factors: first, the halos selected in Websky tend to be more massive than those hosting CMASS galaxies~\citep{White2011}; second, some CMASS galaxies may not be central galaxies or may be offset from the halo center, which will lower the $\bar{\tau}$ profile amplitude~\cite{Bigwood2025}.

As our main focus is to study the extent to which baryonic feedback effects push gas out of the halo, we focus on the shape information of the $\bar{\tau}$ profiles and renormalize the Websky $\bar{\tau}$ profile to match the ACT+CMASS measurement at $\theta_{\rm AP} = 6.0$ arcmin. This accounts for amplitude differences arising from halo mass discrepancies between the simulation and observations, and is shown by the solid lines in Figure~\ref{fig:tau_profile_all}. We exclude points at $\theta_{\rm AP} > 6.0$ arcmin in this and the following comparisons due to their large uncertainties.

We also construct a theoretical $\bar{\tau}$ profile based on the NFW profile and the Websky halo catalog (see Appendix~\ref{app:NFW}), shown as the dashed lines in Figure~\ref{fig:tau_profile_all}. This model represents the distribution expected from dark matter alone, which differs significantly from the gas distribution affected by baryonic feedback processes. This profile is also renormalized to the observed $\bar{\tau}$ profile at $\theta_{\rm AP}=6$ arcmin, and is shown by the dotted lines in Figure~\ref{fig:tau_profile_all}. 

Qualitatively, we see a clear difference between the observed $\bar{\tau}$ profiles and the NFW one. The NFW profile flattens at around $1\,{\rm Mpc}/h$, which is the scale of a typical halo virial radius. This is consistent with the fact that most halo mass lies within the virial radius. In contrast, the observed $\bar{\tau}$ profiles indicate that gas is pushed to 2–3 times the halo virial radius by baryonic feedback effects. The comparison between the Websky $\bar{\tau}$ profiles and the observed ones also shows moderate differences. The Websky profiles flatten somewhat faster than the observed ones, suggesting that baryonic feedback effects in the real universe may be stronger than those in the hydrodynamic simulations~\citep{Battaglia2010} on which the Websky simulation relies to model its gas distribution.

Quantitatively, we measure the SNR of the observational $\bar{\tau}$ profiles at both frequencies with respect to the null detection ($\rm SNR_{null}$), the normalized Websky results ($\rm SNR_{Websky}$), and the normalized NFW profiles ($\rm SNR_{NFW}$). The detailed definition of SNR is presented in Appendix~\ref{app:SNR}, and the results are shown in Table~\ref{tab:SN}.

The measured $\bar{\tau}$ profiles reject the no-kSZ hypothesis at $8.7\sigma$ and $7.3\sigma$ for the f150 and f090 frequencies, respectively. Compared to the Websky simulation, which includes baryonic feedback, the $\bar{\tau}$ profiles of ACT+CMASS show discrepancies with significances of $\rm SNR_{Websky} = 7.4$ for f150 and $6.9$ for f090. The trend of the green (or purple) solid lines relative to the red (or blue) data points in Figure~\ref{fig:tau_profile_all} suggests that the Websky optical depth profile is more concentrated than observed, implying that baryonic feedback in the real Universe is stronger than that implemented in the Websky simulation~\cite{Battaglia2010,Stein2020}. This conclusion is consistent with previous studies which targeted comparisons with various state-of-the-art hydro-simulations~\cite{Schaan2021,Amodeo2021,Hadzhiyska2025,Guachalla2025,Bigwood2024,McCarthy2025,Bigwood2025,Siegel2025,Roper2025}

\subsection{Joint-fitting analysis}
\label{subsec:joint_analysis}

When conducting the joint-fitting analysis, the optical depth $\bar{\tau}$ is simultaneously fitted with the cosmological and nuisance parameters. We also vary the aperture size $\theta_{\rm AP}$ to carry out multiple fitting processes so as to measure the integrated $\bar{\tau}$ profile as a function of $\theta_{\rm AP}$. The results are presented as the open squares with error bars in Figure~\ref{fig:tau_profile_mcmc}. The $\bar{\tau}$ estimates from the joint-fitting analysis and the separate-fitting analysis approach are consistent within $1\sigma$ uncertainties, verifying the reliability of the later.

Although the joint-fitting analysis is more statistically rigorous, the inclusion of additional free parameters dilutes the constraining power on $\bar{\tau}$, leading to larger uncertainties and a lower SNR compared to the separate-fitting analysis. Therefore the main conclusions of this work are presented by the separate-fitting analysis.

\section{Discussion and Conclusion}
\label{sec:conclusion}
We perform in this work a joint analysis of the galaxy power spectrum multipoles and the pairwise kSZ dipole to break the degeneracy between the mean optical depth $\bar{\tau}$ and nuisance parameters, enabling a measurement of the $\bar{\tau}$ profile for the CMASS galaxy sample using the ACT DR6 CMB map. By varying the aperture radius $\theta_{\rm AP}$, our approach indirectly probes the baryon distribution within and around halos. Among these $\theta_{\rm AP}$s, both f150 and f090 achieve their highest SNRs at $\theta_{\rm AP}=2.0\ \rm arcmin$, with values of 7.2 and 6.4, respectively. The $\rm SNR_{null}$s of the $\bar{\tau}$ profile with $\theta_{\rm AP} \leq 6\ \rm arcmin$ against the no-kSZ hypothesis are 8.7 and 7.3 for the f150 and f090 frequencies, respectively.

To ensure consistency between observational and simulated results, we apply the same analysis pipeline to both observational data and the Websky simulation to measure the kSZ signal. The resulting profile shape discrepancies between data and simulation are $\rm SNR_{Websky} = 7.4$ for the f150 band and $6.9$ for the f090 band, suggesting that baryonic feedback in the real universe is stronger than that implemented in Websky simulation. This finding aligns with previous studies that favor stronger feedback scenarios in simulations~\citep{Amodeo2021,Bigwood2024,McCarthy2025,Guachalla2025,Popik2025}. Additionally, the observed $\bar{\tau}$ profile deviates significantly from the NFW prediction, with discrepancies of $\rm SNR_{NFW} = 27.5$ for f150 and $21.3$ for f090. This indicates that dark matter is more centrally concentrated than baryons—consistent with the results of~\citep{Schaan2021}—as baryonic feedback ejects gas from the central regions of halos.

Although the identical methodology we apply to both observational  and simulational data minimizes the risk of introducing unknown systematic biases that could arise from using different analysis techniques, several effects may still contribute to the observed differences between the data and the simulation, and should be carefully considered when interpreting the results. First, the off-centering of galaxies relative to halo centers can lead to a suppression of the kSZ signal at small angular scales, effectively modifying the inferred $\bar{\tau}$ profile. Second, there is an inherent mismatch between the observed galaxy sample and the simulated halo sample, as the latter is selected based on a simple mass threshold rather than a realistic galaxy–halo connection. Since halos of different masses exhibit not only different amplitudes but also different profile shapes~\cite{Siegel2025}, this mismatch may introduce biases in the inferred $\bar{\tau}$ profile.

Importantly, both of these effects can alter the shape of the optical depth profile, leading to a degeneracy with the impact of baryonic feedback. Therefore, while the observed profile being more extended than that in Websky may suggest more efficient redistribution of gas in the real Universe, a definitive interpretation requires improved modeling of these systematic effects. Future work incorporating more realistic galaxy–halo connections, for example through halo occupation distribution modeling combined with lensing constraints~\cite{Hadzhiyska2025b}, will help to further clarify this issue.


{\it Acknowledgments:} Y.Z. acknowledges the support from the National SKA Program of China (2025SKA0150104) 
and the National Natural Science Foundation of China (NSFC) through grant 12203107.

\bibliography{mybib}{}

\appendix
\section{NFW profile}
\label{app:NFW}
We use the dark matter distribution as a tracer of the baryon distribution within the halo to examine the difference between simulation and observation. \citep{Sugiyama2018} provides a prediction for the optical depth $\tau$ of a single halo.

We model the halo profile using the truncated three-dimensional NFW profile, given by
\begin{eqnarray}
\label{eq:NFW_profile}
\rho(r) = \left\{
\begin{array}{cc}
\frac{\rho_s}{(r/r_s)(1+r/r_s)^2}\,, & r < R_{200m} \\[1em]
0\,, & r \geq R_{200m}
\end{array}
\right.
\end{eqnarray}
with
\begin{equation}
\label{rho_s}
    \rho_s = \frac{200}{3} \, \rho_m(z) \, \frac{c_{200m}^3}{\ln(1+c_{200m}) - c_{200m}/(1+c_{200m})}\, .
\end{equation}
The halo mass enclosed within $R_{200m}$, defined as 200 times the mean matter density, satisfies $M_{200m} = 200 \times (4\pi/3) R_{200m}^3 \, \rho_m(z)$, where $\rho_m(z) = \Omega_m(z) \, \rho_{\rm crit}(z)$. The concentration parameter $c_{200m} = R_{200m}/r_s$ is taken from~\citep{Bhattacharya2013}.

The two-dimensional surface density is obtained by integrating along the line of sight:
\begin{eqnarray}
\label{Sigma}
    \Sigma(R) &=& \int_{-\infty}^{+\infty} \rho\!\left(\sqrt{R^2 + z^2}\right) dz\,, \\  
    \Sigma(\theta) &=& \int_{-\infty}^{+\infty} \rho\!\left(\sqrt{\theta^2 / D_A^2 + z^2}\right) dz\,,
\end{eqnarray}
where $\theta = R/D_A$, and $D_A$ is the angular diameter distance to the halo. The mean surface density averaged over the halo sample is
\begin{equation}
\label{Sigma_mean}
    \bar{\Sigma}(\theta) = \frac{\sum_i w_i \, \Sigma_i(\theta)}{\sum_i w_i},
\end{equation}
with $w_i$ denoting the FKP weight of the $i$th halo.

The theoretical optical depth profile is then given by
\begin{equation}
\label{tau_th}
    \tau = \frac{\sigma_{\rm T} \, f_{\rm gas}}{\mu_{\rm e} \, m_{\rm p}} \int \frac{d\boldsymbol{\ell}}{(2\pi)^2} \, W_{\rm AP}(\ell \theta_{\rm AP}) \, \bar{\Sigma}(\boldsymbol{\ell}) \, B(\boldsymbol{\ell}),
\end{equation}
where $W_{\rm AP}$ is the aperture photometry (AP) filter:
\begin{equation}
\label{W_AP}
    W_{\rm AP}(x) = \frac{2}{x} \left[ 2J_1(x) - \sqrt{2} J_1(\sqrt{2}x) \right],
\end{equation}
with $x = \ell \theta_{\rm AP}$. Here, $\sigma_{\rm T}$ is the Thomson scattering cross-section, $\mu_{\rm e} = 1.17$ is the mean particle weight per electron, $m_{\rm p}$ is the proton mass, and $f_{\rm gas} = 0.157$ is the cosmic baryon fraction. The quantities $\bar{\Sigma}(\boldsymbol{\ell})$ and $B(\boldsymbol{\ell})$ are the Fourier transforms of the mean surface density and the Gaussian beam, respectively. The beam function is $B(\ell) = \exp(-\sigma_B^2 \ell^2 / 2)$, with $\sigma_B = \mathrm{FWHM} / \sqrt{8\ln 2} = 0.4247 \, \mathrm{FWHM}$.

The redshift distribution of halos used in the NFW model is taken from the Websky simulation, matched to the CMASS sample. The beam FWHM is set to match the ACT maps at the f150 and f090 frequencies. This allows us to consistently compare the NFW-based prediction with both the simulation and the observational results.

\section{Estimation of signal-to-noise ratio}
\label{app:SNR}

We define the SNR with respect to the null-detection hypothesis as \(\rm SNR_{null}\), with respect to the Websky simulation as \(\rm SNR_{Websky}\), and with respect to the NFW model as \(\rm SNR_{NFW}\). All data points at $\theta_{\rm AP} > 6.0$ arcmin are excluded from the SNR calculations due to their large uncertainties. Here, \(\rm SNR_{null}\) quantifies the kSZ signal of the \(\bar{\tau}\) profiles, while \(\rm SNR_{Websky}\) and \(\rm SNR_{NFW}\) respectively measure the differences between the ACT+CMASS profiles and the Websky or NFW model predictions. The SNRs are estimated as follows:
\begin{equation}
\label{Delta}
    {\rm SNR}_{\rm model} \equiv \sqrt{\chi^2}\, ,
\end{equation}
where
\begin{equation}
\label{eq:chi2}
    \chi^2 = ({\rm data} - {\rm model})^\top \, {\rm Cov}_{\rm model}^{-1} \, ({\rm data} - {\rm model})\, .
\end{equation}
Here, "data" represents the $\bar{\tau}$ profiles of the ACT+CMASS measurements, including those for f150 and f090, shown as the red and blue points, respectively, in Figure~\ref{fig:tau_profile_all}. The "model" refers to the theoretical model of the $\bar{\tau}$ profiles with respect to the "null", "Websky", and "NFW" cases, respectively. ${\rm Cov}_{\rm model}$ is the covariance matrix that describes the correlations among the different $\bar{\tau}$ measurements.

For $\rm SNR_{null}$, "model" is the zero vector ($\boldsymbol{0}$), representing the null hypothesis, and its covariance matrix ${\rm Cov}_{\rm null}$ is estimated using Equation~(\ref{eq:Cij_tau}) applied to the ACT+CMASS measurements. The corresponding correlation matrix is shown in the right panel of Figure~\ref{fig:tau_profile}.

For $\rm SNR_{Websky}$, "model" corresponds to the re-normalized $\bar{\tau}$ profiles of the Websky simulation, shown as solid lines in Figure~\ref{fig:tau_profile_all}. Its covariance matrix ${\rm Cov}_{\rm Websky}$ differs slightly from ${\rm Cov}_{\rm null}$. To simplify the incorporation of the Websky measurement uncertainty, we keep the correlation coefficients of ${\rm Cov}_{\rm Websky}$ the same as those of ${\rm Cov}_{\rm null}$ and only rescale the diagonal elements. The covariance matrix of the Websky $\bar{\tau}$ profile is first estimated using Equation~(\ref{eq:Cij_tau}); we then re-normalize its diagonal elements by the square of the re-normalization factor of the Websky measurements. The diagonal elements of ${\rm Cov}_{\rm Websky}$ are finally taken as the sum of the diagonal elements of the ACT covariance matrix and these re-normalized Websky diagonal elements, thereby accounting for the measurement uncertainty of the Websky simulation.

For $\rm SNR_{NFW}$, "model" is the re-normalized $\bar{\tau}$ profiles of the NFW model, shown as dotted lines in Figure~\ref{fig:tau_profile_all}, and the covariance matrix is the same as that used for $\rm SNR_{null}$.

\end{document}